\documentclass[12pt,letterpaper]{article}
\usepackage{graphics}
\usepackage{amssymb,epsfig,amsmath,euscript,array}
\usepackage{subfigure}
\usepackage{color}
\usepackage{slashed}
\usepackage{axodraw}

\newcommand{\mix}{\chi}

\newcommand{\0}{\mathrm{L}}
\newcommand{\T}{\mathrm{T}}
\newcommand{\OP}{\omega_\mathrm{P}}
\newcommand{\muu}{ m_{\gamma^\prime} }

\begin{document}


\hfill{DESY 07-211}\\

\vspace{20pt}

\begin{center}

{\Large \bf Helioscope Bounds on Hidden Sector Photons} \\[1.5ex]

\vspace{30pt}

{Javier~Redondo}

\vspace{7pt}

{\small \em
{}$^4$Deutsches Elektronen-Synchrotron DESY,\\
Notkestrasse 85, D-22607  Hamburg, Germany}

\vspace{40pt}

{\bf Abstract}
\end{center}
The flux of hypothetical ``hidden photons" from the Sun is computed under the assumption that they interact with normal matter only through kinetic mixing with the ordinary standard model photon.  Requiring that the exotic luminosity is smaller than the standard photon luminosity provides limits for the mixing parameter down to $\mix\lesssim 10^{-14}$, depending on the hidden photon mass. Furthermore, it is pointed out that helioscopes looking for solar axions are also very sensitive to hidden photons. The recent results of the CAST collaboration are used to further constrain the mixing parameter $\mix$ at low masses ($\muu<1$ eV) where the luminosity bound is weaker. In this regime the solar hidden photon flux has a sizable contribution of longitudinally polarized hidden photons of low energy which are invisible for current helioscopes.

\noindent

\setcounter{page}{0}
\thispagestyle{empty}
\newpage
\section{Introduction}

The standard model (SM) provides an accurate description of particle physics below the electroweak scale but it is generally though not to be valid up to arbitrary energies. Extensions of this celebrated scheme, invoked to cure diseases like the strong CP, hierarchy or flavor problems, often involve higher gauge symmetries and further matter content. Moreover, string theory is a preferred candidate for the unification of quantum mechanics and general relativity where additional gauge and matter fields are assured. 
At low energies, some of these new fields can arrange into a ``hidden sector" if only very massive particles (or gravity) mediate interactions between them and the SM ``visible sector".

Of course, depending on the scalar content of the theory, gauge symmetries can either be spontaneously broken or remain exact. Then, the corresponding ``hidden" bosons could have in principle an arbitrary mass. If this is small enough, these hidden bosons can have a very rich phenomenology at present affordable energy scales. 

The simplest case concerns just a novel $U(1)_\mathrm{h}$ symmetry and its corresponding gauge boson, henceforth called ``hidden" photon. The interplay between this hidden photon and the SM photon modifies the predictions of quantum electrodynamics \cite{Okun:1982xi}, often claimed to be the most accurate of all physical theories so far, thus constraining the hidden photon parameters. We can turn this argument in the opposite direction: the constraints on hidden photon parameters give us information about how accurate is the QED description of nature at low energies.

A number of laboratory experiments has been devoted to the search of hidden photons, the resulting bounds being strongly dependent on the hidden photon mass. For masses corresponding to macroscopic length scales, experiments testing the Coulomb law \cite{Williams:1971ms,Bartlett:1988yy} set strong constraints on hidden photons, but still they could be largely improved by experiments dealing with high quality microwave cavities \cite{Jaeckel:2007ch}. In the microscopic range, laser experiments are also becoming very powerful probes of hidden sector particles \cite{Cameron:1993mr, Gies:2006ca,Ahlers:2006iz,Ahlers:2007rd,Jaeckel:2007gk,Robilliard:2007bq,Chou:2007zz,Ahlers:2007qf}. At atomic distances, comparison of the Rydberg constant for different atomic levels gives interesting but weak bounds \cite{Beausoleil:1987,Garreau:1987}.
Finally, particle colliders extend the mass range until typical electroweak scales \cite{Kors:2004dx,Feldman:2006ce,Chang:2006fp,Kumar:2006gm,Feldman:2007nf}.

On top of that, the evolution of stars turns out to be the most sensitive ``laboratory" to study properties of novel low mass weakly interacting particles \cite{Raffelt:1996wa,Raffelt:1999tx}. Even with tiny couplings to electrons and protons, they might be still copiously created in the interior of hot and dense stars. Because of their weak interactions they might abandon the star without further scattering, accelerating the consumption of nuclear fuel, and therefore the stellar evolution \cite{Dicus:1978fp,Frieman:1987ui}. Our present observational data on stellar evolution can strongly constrain this novel luminosity, although the bounds can be relaxed in some concrete models \cite{Masso:2005ym,Masso:2006gc,Jaeckel:2006id,Jaeckel:2006xm,Brax:2007ak}.

Interestingly enough, in this case the Sun itself could be a copious emitter of weakly interacting particles, that could eventually be detected at Earth inside a sensitive detector \cite{Sikivie:1983ip} (as this is actually the case with neutrinos). Several of these so-called ``helioscopes" \cite{vanBibber:1988ge} have been built with the aim of detecting solar axions \cite{Lazarus:1992ry,Moriyama:1998kd,Zioutas:2004hi} and remarkably, the CAST collaboration has even recently surpassed the sensitivity of the energy loss arguments for the axion coupling to two photons \cite{Andriamonje:2007ew}. As we will see, helioscopes can also detect hidden photons so the CAST limits can also be used to constrain the solar hidden photon flux.

The energy loss argument and helioscope bounds for hidden photons were already studied in  \cite{popov:1991,Popov1999}. However, in this paper the author does not consider either the possibility of a resonant production, which can enhance enormously the hidden photon flux, nor the emission of transversely polarized hidden photons.

In this paper we compute the energy loss bounds using the latest solar data \cite{Bahcall:2004pz} and derive the CAST helioscope bounds. In particular, it is shown that the new CAST results are extremely sensitive to solar hidden photons, providing the strongest constraint of their existence in the mass range $\muu\sim 0.01-1$ eV. Moreover, accounting for the resonant production improves the energy loss bounds in \cite{Popov1999} roughly up to 1 order of magnitude. 

The paper is organized as follows: In Sec.\ref{Sproduction} the hidden photon solar emission is derived while the principles of the CAST helioscope detection are reviewed in Sec.\ref{Shelioscope}. We compute the energy loss and CAST bounds in Sec.\ref{Sbounds} and finally the conclusions are presented.

\section{Hidden photon production in the Sun}\label{Sproduction}

At low energies, the dynamics of the two photon system can be described by means of the following Lagrangian, 
\begin{equation}
{\cal L} = {\cal L}_{\mathrm{MX}}+{\cal L}_{\mathrm{h}}+{\cal L}_{\mathrm{\chi}}=-\frac{1}{4}A_{\mu\nu}A^{\mu\nu}+A_{\mu}j^{\mu} -\frac{1}{4}B_{\mu\nu}B^{\mu\nu}+\frac{1}{2}\muu^2B_\mu B^\mu -\frac{1}{2}\chi A_{\mu\nu}B^{\mu\nu}
\end{equation}
which is the sum of the Maxwell Lagrangian for the standard model photon field $A^\mu$ with its corresponding source $j^\mu$ (the electric current of electrons and protons), the Proca Lagrangian for the massive hidden photon $B^\mu$ and a gauge invariant mixing term. This term has been discussed in some detail in the literature in a variety of contexts \cite{Holdom:1985ag,Holdom:1986eq,Holdom:1990xp,Dienes:1996zr,Lust:2003ky,Abel:2003ue,Blumenhagen:2006ux,Abel:2006qt,Babu:1997st,Foot:1991kb}. Although $\chi$ is a priori a free parameter that could be even of order 1, the experimental evidence indicates that it should be much smaller. A natural explanation for this is that while $\chi$ could be zero because of symmetry reasons at a high energy scale, the integration of high energy quantum fluctuations will end up inevitable in a small nonzero value in the low energy theory. Typical values comprised in the literature range from $10^{-16}$ to $10^{-2}$ \cite{Dienes:1996zr}. The lack of a mass mixing term is of course due to the protection of the $U(1)_\mathrm{em}$ symmetry, which in principle we would like to preserve. We do not have to write a tree level coupling $B_\mu j^\mu$ because it can be eliminated (if small) by a harmless redefinition of the photon field and therefore absorbed in $\mix$. This could not be done if the $B^\mu$ field couples with different strength to electrons and protons. Such a situation, however, is severely constrained from experiments testing the neutrality of matter \cite{Marinelli:1983nd} and arguments concerning the existence of leptonic and baryonic forces \cite{Grifols:1988fv} and therefore we will not consider it.

The presence of a mixing term simply  states the fact that the $A^\mu$ and $B^\mu$ fields are non-orthogonal. This can be easily seen if we notice that the Feynman rule for the $A_{\mu\nu}B^{\mu\nu}$ vertex is simply the inverse (non diagonal) massless propagator. In order to gain further understanding of the physics to be discussed it is highly recommendable to express the Lagrangian in terms of the state that is orthogonal to the photon, and therefore \emph{sterile} with respect to local electromagnetic interactions,  defined as
\begin{equation}
S^\mu = B^\mu+\mix A^\mu \ .
\end{equation}
This invites us to renormalize the electric charge $\sqrt{1-\mix^2}A^\mu\rightarrow A^\mu$ and generates non-diagonal mass terms, 
\begin{equation}
{\cal L}_m= \frac{1}{2}\muu^2 S^\mu S_\mu-\mix\muu^2 S^\mu A_{\mu}+\frac{1}{2}\mix^2\muu^2 A^\mu A_{\mu} \ , 
\end{equation}
which indicate that photons oscillate into sterile states during free propagation, like in the case of neutral kaons or neutrinos.

The equations of motion (EOM) in the Lorentz gauge for such a system are, in Fourier space, 
\begin{eqnarray}
(K^2g^{\mu\nu} - \Pi^{\mu\nu}(K)-\mix^2\muu^2) A_\nu(K)+ \chi \muu^2 S^\mu(K)&=&0 \ \ , \nonumber  \\
(K^2-\muu^2) S^\mu(K)+ \chi\muu^2 A^\mu(K)&=&0 \ , 
\end{eqnarray}
where $K$ is the four momentum (from now on I will skip writing it) and I have included the effects of the photon interactions with the medium via the polarization tensor $\Pi^{\mu\nu}$, which has a clear physical interpretation as the photon self energy in the medium\footnote{In this discussion I follow the exposition and notation of \cite{Raffelt:1996wa}.}. In a homogeneous, isotropic and CP conserving medium it has only two independent components ,$\pi_\T$ and $\pi_\0$, corresponding to two transverse ($\bf{k} \cdot \bf{A}=0$) and one longitudinal (Langmuir waves with $\bf{k} \times \bf{A}=0$) electromagnetic excitations, which in a plasma are called \emph{plasmons}. Taking advantage of the isotropy, we chose to focus on propagation along the $z$-direction $K=(\omega;0,0,k)$ and we can write a simple set of orthogonal polarization vectors\footnote{We will not consider space-like plasmons in this paper so the denominators will cause no trouble.} as
\begin{equation}
\epsilon_{T_1}^\mu (0;1,0,0) \ \ ; \ \ 
\epsilon_{T_2}^\mu (0;0,1,0) \ \ ; \ \
\epsilon_\0^\mu \frac{1}{\sqrt{\omega^2-k^2}}(k;0,0,\omega) \ \ ,
\end{equation}
satisfying $\epsilon^\mu_a\epsilon^*_{b\mu}=-\delta_{ab}$ and define
\begin{equation}
\Pi^{\mu\nu}(K)=-\sum_a \epsilon_a^\mu \epsilon^{*\nu}_a \pi_a(K) \ .
\end{equation}
Thus we can write the EOM for the different components $A_a=\epsilon_a^\mu A_\mu$ (idem for $S_a$) as
\begin{eqnarray}
(\omega^2-k^2 - \pi_{a}-\mix^2\muu^2) A_a+ \chi\muu^2 S_a&=&0 \ \ , \nonumber \\ 
(\omega^2-k^2 - \muu^2)  S_a+ \chi \muu^2 A_a&=&0  \ \ .      \label{eom1}
\end{eqnarray}
The evolution of T and L plasmons is decoupled. 

The lowest order contribution to $\Pi^{\mu\nu}$ in  a plasma comes from coherent forward scattering off the thermal bath of electrons \cite{Altherr:1992mf,Braaten:1993jw},  providing refractive (real) parts for $\pi_{\T,\0}$. For typical solar plasmon energies, $\lesssim$ 10keV$ \ll m_e$ we can neglect the electron velocity dispersion in our calculations getting \cite{Braaten:1993jw}
\begin{equation}
\mathrm{Re} \{\pi_\T\} \equiv m_\T^2 \simeq \OP^2
\hspace{1cm};\hspace{1cm}
\mathrm{Re} \{\pi_\0 \} \equiv m_\0^2 \simeq \OP^2-k^2
\label{RE} \ ,
\end{equation}
with the plasma frequency $\OP$ given by
\begin{equation}
\OP^2 \simeq \frac{4\pi\alpha}{m_e} n_e
\ \ , \label{OP}
\end{equation}
where $n_e$ is the electron number density. 

Leaving aside the effects of the hidden sector photons, the solutions from the equations of motion eq.~\eqref{eom1} tell us that T-plasmons behave as massive particles with an ``effective mass"" given by $\OP$
 but L-plasmons oscillate at a frequency $\omega\sim\OP$, almost independent of the wave number $k$ (the group velocity is suppressed by a small factor $\sim 3T/m_e$). Then, while T-plasmons are always time-like, L-plasmons can also be light or space-like. Explicitly one finds 
\begin{equation}
\omega^2 \simeq \OP^2+k^2 \hspace{1cm}  (\mathrm{T})  \hspace{0.2cm} ; \hspace{1cm} \omega^2 \simeq \OP^2+\frac{3T}{m_e}k^2 \hspace{1cm}  (\mathrm{L}) \hspace{0.5cm} ,\label{dispersion}
\end{equation}
Where I have kept the lowest order terms in $k$. 

In order to get the lowest nonzero contributions to the imaginary parts of $\pi_{\T,\0}$ we need to include photon absorption and dispersion. Following Weldon \cite{Weldon:1983jn}, the imaginary part of the photon self energy is proportional to the difference of the photon absorption and photon production probabilities ($\Gamma^\mathrm{A,P}$) by means of 
\begin{equation}
\mathrm{Im}\ \pi_a  \equiv -\omega\Gamma_a = -\omega\left(\Gamma_a^\mathrm{A}-\Gamma_a^\mathrm{P}\right)=-\omega(1-e^{-\frac{\omega}{T}})\Gamma^\mathrm{A}_a    \label{weldon} \ , 
\end{equation}
where in the last equality thermodynamic equilibrium\footnote{Even in the absence of thermodynamic equilibrium the result holds because of unitarity \cite{Weinberg:1979bt,Weldon:1983jn}. } is assumed, implying
\begin{equation}
\Gamma^\mathrm{P}=\Gamma^\mathrm{A}e^{-\frac{\omega}{T}} \ \ . \label{LTE}
\end{equation}
The result eq. \eqref{weldon} reminds on an absorption rate corrected for stimulated emission.

The dominant source of opacity in the Sun is inverse bremsstrahlung (also called free-free absorption) which is very efficient at low and intermediate energies. Compton scattering provides a smaller but energy independent contribution (given that $\omega\ll m_e$) which is crucial for the highest energies. For transversely polarized photons of energy sufficiently above the plasma frequency we get
\begin{equation}
\Gamma_\T= \frac{16\pi^2\alpha^3}{3 m_e^2\omega^3}\sqrt{\frac{2\pi m_e}{3 T}}n_e \sum_i  Z^2_in_i\overline{g}_{ff,i}(1-e^{-\frac{\omega}{T}}) +\frac{8\pi\alpha^2}{3 m_e^2}n_e  \ , \label{IMT}
\end{equation}
where $n_i$ is the number density of ions of charge $eZ_i$ -which as a good approximation we will take to be only Hydrogen and Helium- and $\overline{g}_{ff,i}$ is the Boltzman averaged Gaunt factor\footnote{For a review on the subject see \cite{Brussaard1962}. In the numerical estimates of this paper I have recalculated the Gaunt factor using the exact Sommerfeld-Maue formula \cite{Sommerfeld1935} and a Laguerre-Gaus quadrature as proposed in \cite{Grant1958}.} which accounts for the deviations from the classical expression derived by Kramers \cite{Kramers1923}. A more accurate calculation should include the effects of free-bound and bound-bound transitions, which might acquire some relevance at the relative low temperatures of the solar external layers. Note that in eq.~\eqref{IMT} we are also assuming that all the particle species are completely ionized. Again this approximation will fail at the solar external layers and corrections should be included by solving the Saha equation. We will refer to the validity of these simplifications later on.

For $\pi_\0$ the situation is more complicated since formulas for the interactions of longitudinal plasmons are difficult to derive or to find in the literature. However it turns out that in the most interesting case, where L-plasmons are time-like, the bulk emission is not sensitive to the details of $\Gamma_\0$. For completeness, however,  we can write the simplest contribution, Thomson dispersion (derived in Appendix \ref{A}), giving
\begin{equation}
\Gamma_\0= \frac{8\pi\alpha^2}{9 m_e T}\frac{k}{\omega} n_e  \label{IML} \ .
\end{equation} 

\begin{figure}[t]
\begin{center}
\includegraphics[width=6.5cm]{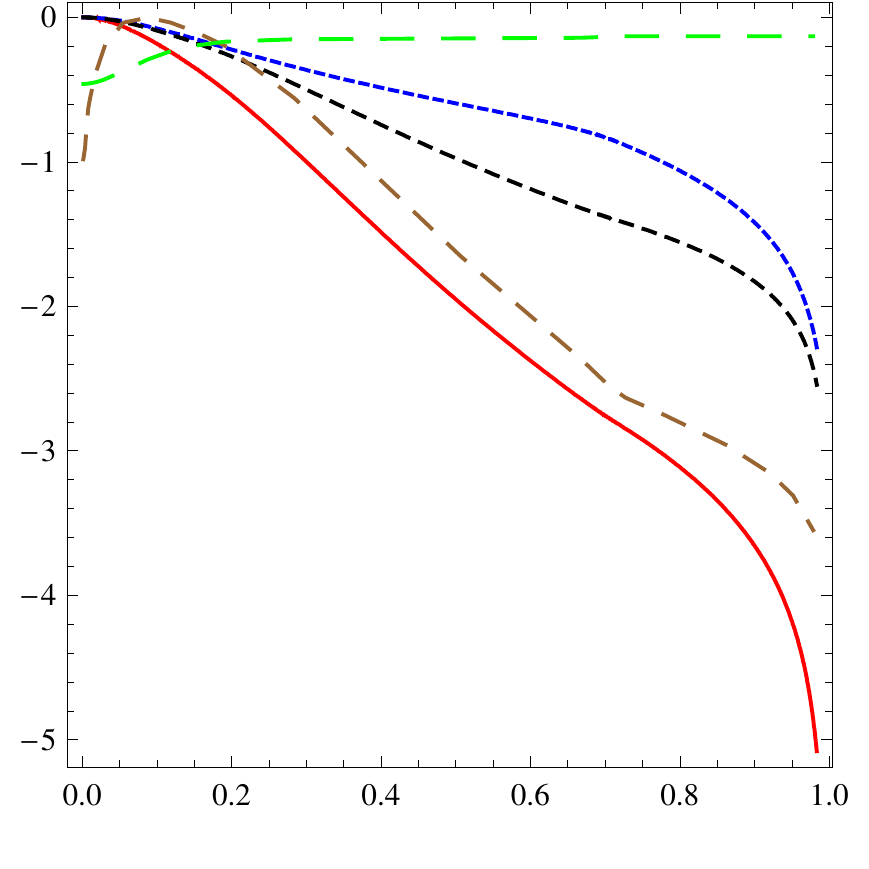}
\includegraphics[width=6.5cm]{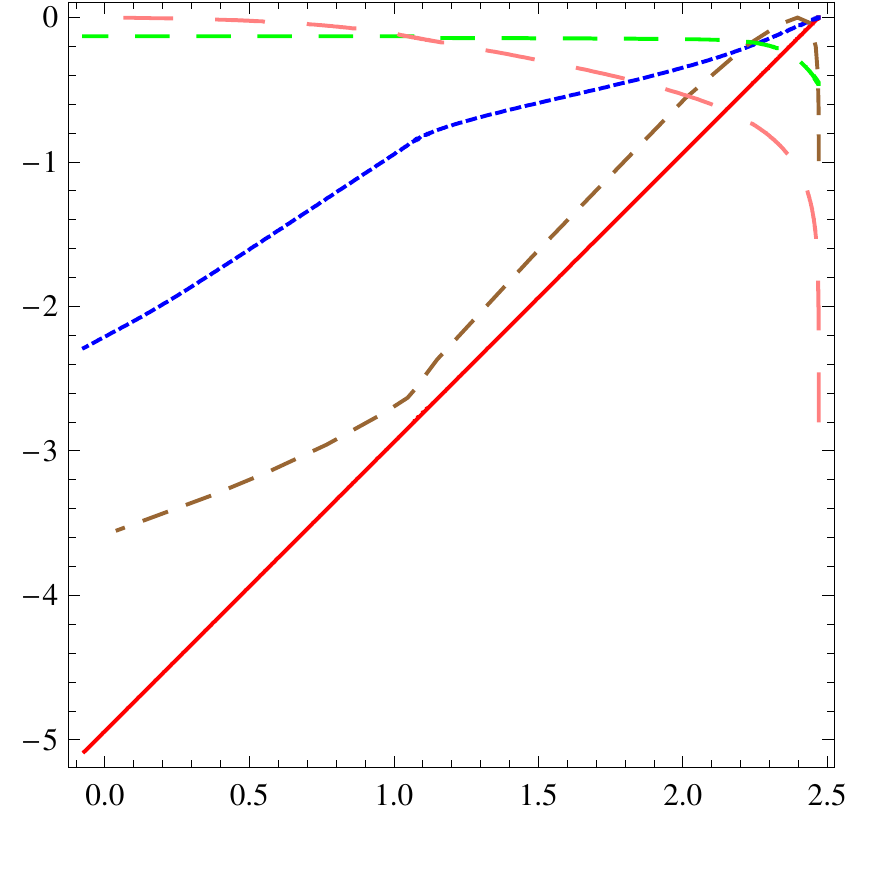}
\Text(-277,1)[c]{\scalebox{1.2}[1.2]{$r$}}
\Text(-90,1)[c]{\scalebox{1}[1]{Log$_{10}\ \OP\left[\mathrm{eV}\right]$}}
\vspace{.5cm}
\caption{Values of the solar parameters relevant for this work plotted as a function of the normalized solar radial coordinate $r$ (left) and the plasma frequency $\OP$ (right) in decimal logarithmic scale. From small to large dashing these are the electron density $n_e$ (red), temperature $T$ (blue), plasma frequency (black), $d\OP^2/dr$ (brown), Hydrogen mass fraction $X$ (green) and radial coordinate $r$ (pink). Except for $X$, they are normalized to their largest values $6.07\times 10^{25}$ cm$^{-3}$, $1350$ eV,  $295.5$ eV, $5.8\times 10^{-4}$ eV$^2$ m$^{-1}$ and $R_\odot=6.96\times 10^8$ m, respectively (taken from \cite{Bahcall:2004pz}).}
\label{solarmodel}
\end{center}
\end{figure}

The solar model BP05(OP) \cite{Bahcall:2004pz} provides the latest available data on the solar interior compatible with helioseismology and neutrino fluxes. Magnitudes like temperature ($T$), mass density ($\rho$) and the mass fraction of the most important atomic elements ($X_i$) are tabulated as a function of the solar radial coordinate $r$. Assuming that the plasma is locally neutral we can compute the electron density as the sum over species of the number of protons through $n_e=\rho/m_u \sum_i Z_i X_i  / A_i$, with $Z_i,A_i$ the atomic and mass number of species $i$ and $m_u$ the atomic mass unit. The relevant parameters for this work are plotted in Fig.~\ref{solarmodel}. In particular notice that the electron density strongly depends on the position in the solar interior. This introduces an implicit position-dependence in the EOM \eqref{eom1}.

Now that we have learned about the standard plasmon propagation in the Sun it is time to address the effects of mixing with the hidden photon $B$. Assuming slow spatial variation of the electron density, the EOM in eq.~\eqref{eom1} are diagonalized by the following shift 
\begin{equation}
A_a \rightarrow
   \tilde A_a + \mix\frac{\muu^2}{\pi_a-\muu^{2}}\tilde S_a \hspace{.5cm} ; \hspace{.5cm}
S_a\rightarrow    
 \tilde S_a +   \mix\frac{\muu^2}{\muu^2-\pi_a} \tilde A_a
\label{shift}
\end{equation}
as long as the non diagonal elements are much smaller than $1$. I will refer to this condition as \emph{weak mixing} (WM) and justify later that it is satisfied in all the relevant cases. 

The states $\tilde A,\tilde S$ have decoupled evolution and satisfy dispersion relations similar to $A$ and $B$ in the absence of kinetic mixing, 
\begin{eqnarray}
\omega^2-k_{A_a}^2&=&\tilde \pi_a(\omega,k_{A_a}) = \pi_a(\omega,k_{A_a}) + {\cal O}(\chi^2) \ ,   \nonumber \\
\omega^2-k_{S_a}^2&=&\tilde\muu_a^2(\omega,k_{S_a}) = \muu^2 + {\cal O}(\chi^2) \ .
\label{masses}
\end{eqnarray}
Therefore an originally pure-plasmon state $A$ with energy $\omega$ produced in a small region labeled $z=0$ will evolve as a linear combination of the propagating states,
\begin{equation}
A_a(t,z) \simeq  e^{i\omega t-ik_{A_a} z} \tilde A_a+\frac{\mix\muu^{2}}{\pi_a(\omega,k_{S_a})-\muu^{2}
} e^{i\omega t-ik_{S_a} z}\tilde S_a  \ , \label{Ainitial}
\end{equation}
leading in principle to $A-S$ oscillations. However, in the Sun the imaginary part of the plasmon dispersion relation is big enough to damp completely the $\tilde A$ component after a short distance $\sim \Gamma_a^{-1}$. 

The $\tilde S$ component is damped as well since it receives an ${\cal O}(\mix^2)$ imaginary contribution $\propto \Gamma_a$. Once we know what values of $\mix$ we can bound we can go back to this point to show that this absorption is negligible. 

Finally we have to take into account that this solution holds locally in the Sun, or at least in a region where $n_e$ (and therefore $\pi_a$) does not change very much. At every position of the Sun we should define $\tilde S_a(r)$ inverting eq.~\eqref{shift} and using eqs.\eqref{RE}-\eqref{IML} with the function $n_e(r)$ given by the solar model. Interestingly enough, as long as $\Gamma_a$ is sufficiently large, the corresponding $\tilde A_a(r(z))$ component will be quickly absorbed and in practice $A_a(t,z)$ will ``follow" $\tilde S_a(r(z))$ as it travels out of the Sun. 

Eventually, a plasmon will exit the Sun as a $\tilde S_a(R_\odot)$ state, 
\begin{equation}
\tilde S_a(R_\odot) = \tilde S_a(n_e(R_\odot)\rightarrow 0) = B_a \ ,   \hspace{1cm}     
\end{equation}
with a probability
\begin{equation}
P_{A_a\rightarrow B_a} = | \langle S_a | A_a(R_\odot)\rangle |^2 \simeq \mix^2 \frac{\muu^4}{(m^2_{a,0}-\muu^2)^2+(\omega\Gamma_{a,0})^2} \ , \label{PAB}
\end{equation}
where $m^2_{a,0}=m^2_{a}(r_0)$ and $\omega\Gamma_{a,0}=\Gamma_a(r_0)$ are the real and imaginary parts of $\pi_a$ evaluated at the production point, at a distance $r_0$ from the solar center.

Let me remark that the hidden photons exiting the Sun will have the same energy than their original plasmons, \emph{i.e.} of order eV up to $\sim$10 keV. As $B_a$ are propagating states in vacuum they will move away from the solar surface without ``flavor" oscillations. 

The rate of hidden photons exiting the Sun will be then proportional to the plasmon generation rate, i.e. the rate at which plasmons are produced from electron and proton interactions in the plasma. Such a generation rate can be derived from eqs.~\eqref{IMT} and \eqref{IML} by using eq.~\eqref{weldon} and eq.~\eqref{LTE}. We find that this equals\footnote{This is of course related to Kirchhoff's law of thermal radiation.} $\Gamma_a$ defined in eq.~\eqref{weldon} weighted by the Bose-Einstein distribution $n_\mathrm{BE}=n_\mathrm{BE}(\omega,T)=[e^{\frac{\omega}{T}}-1]^{-1}$. This is particularly transparent in the case of scattering since every photon emitted comes from a thermal photon ``absorbed" and the number of these photons is given by the Bose-Einstein function.

We can immediately compute the flux of hidden photons that arrive at the Earth by integrating this rate over the solar model \cite{Bahcall:2004pz}, 
\begin{equation}
d\Phi_a= \frac{1}{4\pi R^2_\oplus} \int_0^{R_\odot} \hspace{-0.2cm}4\pi r^2 d r \frac{k^2dk}{2\pi^2} \frac{\mix^2\muu^4}{(m^2_a-\muu^2)^2+(\omega\Gamma_a)^2} \frac{\Gamma_a}{e^{\frac{\omega}{T}}-1}  \ ,  \label{generalflux} 
\end{equation}
where one has to keep in mind that $m^2_a$ and $\Gamma_a$ depend implicitly on $r$ through $n_e(r)$ and eqs.\eqref{RE}-\eqref{IML} and on $\omega=\omega(k)$ given by eq.~\eqref{dispersion} . $R_\oplus$ is the average Sun-Earth distance $\sim 150 \times10^9$ m. We discuss separately the emission of T and L hidden photons because the different dispersion relations lead to a completely different phenomenology.

\subsection{$B_\T$ production}

Using the dispersion relation in eq.~\eqref{dispersion}(T) and multiplying by a factor of 2 to account for the 2 different T-polarizations, we get
\begin{equation}
\frac{d\Phi_\T}{d\omega} = \frac{1}{4\pi R^2_\oplus} \int_0^{R_\odot} 4\pi r^2 d r \frac{1}{\pi^2} \frac{\omega\sqrt{\omega^2-\muu^2}}{e^{\frac{\omega}{T}}-1}  \frac{\mix^2\muu^4}{(\OP^2-\muu^2)^2+(\omega\Gamma_\T)^2} \Gamma_\T  \ . \label{Tflux} 
\end{equation}
Given that in the solar model $1$ eV $\lesssim\OP\lesssim 295$ eV, this expression has three clearly differentiated regimes that we discuss separately.

\subsubsection{Suppressed production ($\mathbf{\muu\ll 1}$ eV)} 
\begin{figure}[t]
\begin{center}
\includegraphics[width=7cm]{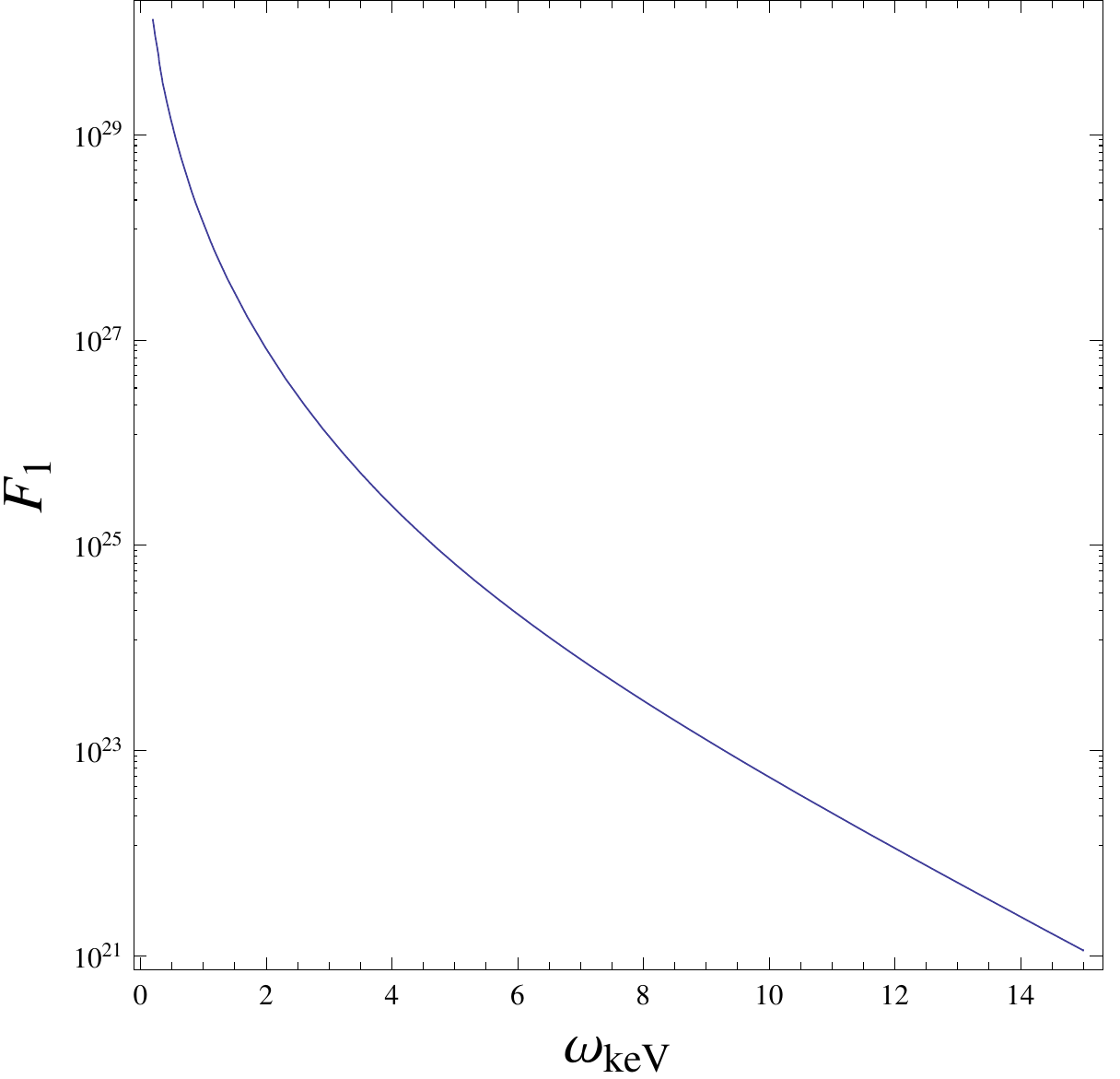}\hspace{1cm}
\includegraphics[width=7cm]{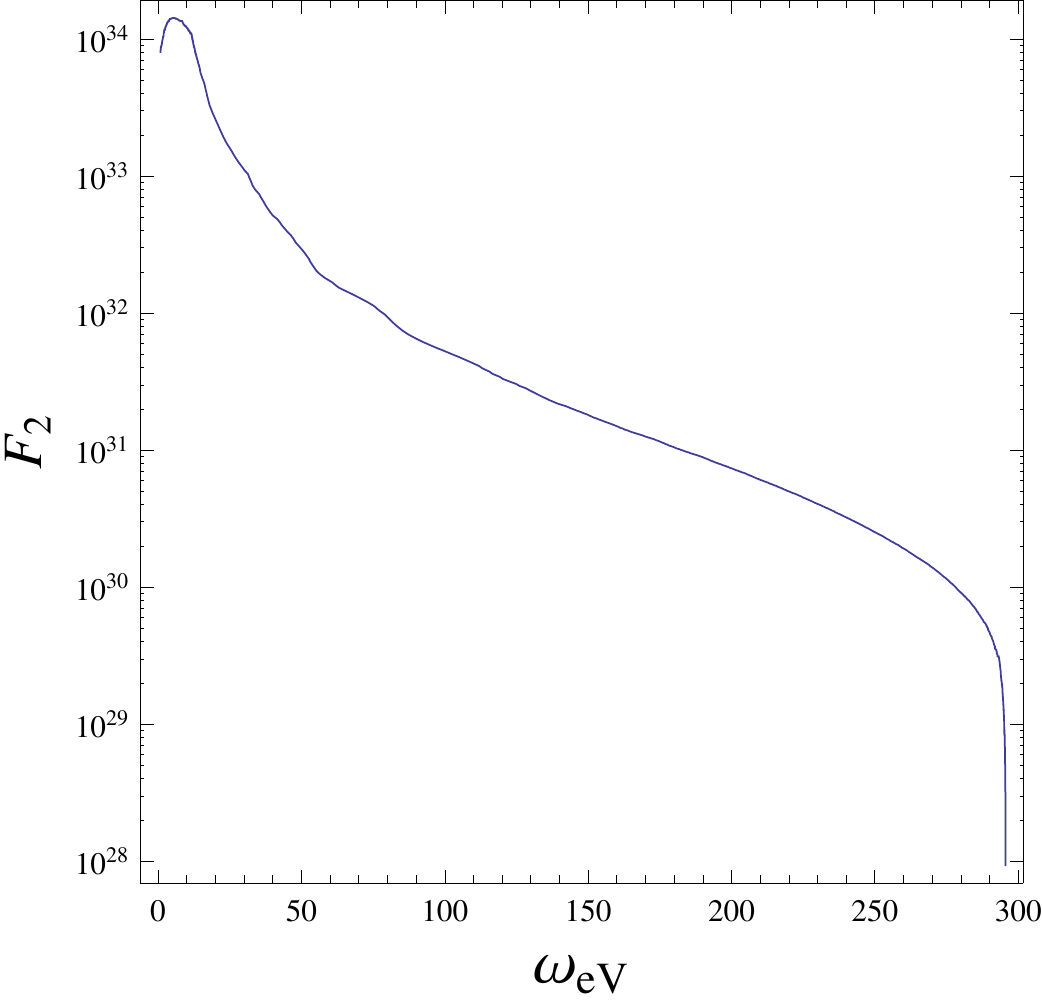}\hspace{1cm}
\caption{The  $F_1$ and $F_2$ functions give the flux of solar transverse and longitudinal $B$'s at the Earth for $\muu\ll1$ eV. Notice the different energy scales, only $eV$ L-hidden photons are emitted while the spectrum of T-modes extends to X-ray energies, although considerably suppressed. See the text for details.}
\label{F}
\end{center}
\end{figure}

For $\muu\ll 1$ eV we can safely neglect the terms involving $\muu$ and $\Gamma_\T$ (typically $\Gamma_\T<\OP$) in the denominator of eq.~\eqref{PAB} and in the square root. Then, all dependence on the hidden photon parameters, $\mix^2\muu^4$, factors out of the integral in eq.~\eqref{Tflux} and we have 
\begin{equation}
\frac{d\Phi}{d\omega} =  \chi^2\muu^4 \frac{1}{4\pi R^2_\oplus} \int_0^{R_\odot} 4\pi r^2 d r \frac{1}{\pi^2} \frac{\omega^2}{e^{\frac{\omega}{T}}-1}  \frac{\Gamma_\T}{\OP^4} \equiv \mix^2\left(\frac{\muu}{\mathrm{eV}}\right)^4\frac{F_1(\omega)}{\mathrm{cm}^2\ \mathrm{s}\ \mathrm{eV}}    \ ,  \label{flux_1} 
\end{equation}
where the dimensionless function $F_1(\omega)$ is plotted in Fig.~\ref{F}. The spectrum peaks at low energies because of two reasons: on the one hand the bremsstrahlung production decreases as $\omega^{-3}$ and on the other hand the production at the solar core, from where the most energetic photons are expected, is suppressed by the largest values of $\OP$. 

As we will see, this suppression regime is the most interesting for helioscopes. A simple analytical formula for $F_1$ can be useful therefore for more delicate future analysis. A fit similar to the axion flux in \cite{Andriamonje:2007ew} has provided, 
\begin{equation}
F_1(\omega)= 2.7 \times 10^{28} E^{-2.98}e^{-\frac{E}{1.4}}\ , 
\end{equation}
where $E=\omega/$keV. The fit reproduces the numerical results between $0.5$ and $5$ keV with 10\% accuracy.

The solar luminosity in hidden sector photons in this case is
\begin{equation}
W_{\gamma'} =  \chi^2\muu^4  \int_0^{R_\odot} 4\pi r^2 d r \frac{1}{\pi^2} \frac{\omega^3}{e^{\frac{\omega}{T}}-1}  \frac{\Gamma_\T}{\OP^4} = 8 \times 10^{43}\ \chi^2 \left(\frac{\muu}{\mathrm{eV}}\right)^4\ \  \mathrm{Watt} \ . \label{W_1}
\end{equation}

It is interesting to point out that for low energies $\sim$ eV most of the production comes from the outer layers of the Sun. In this region, neglecting bound-free and bound-bound transitions or the ionization fraction is not completely justified. However, on the one hand they contribute relatively less than higher energies to the energy loss, and on the other hand the CAST bounds (based on keV energies for which eq.~\eqref{IMT} is accurate) are more restrictive. Therefore, unless this uncertainties imply a huge increase on the energy loss, which is not likely, or an helioscope focuses on these energies, there is no need of correcting eq.~\eqref{IMT}.

\subsubsection{Resonant production ($\mathbf{1}$eV$\mathbf{\lesssim\muu \lesssim295}$ eV)}

\begin{figure}[t]
\begin{center}
\includegraphics[width=10cm]{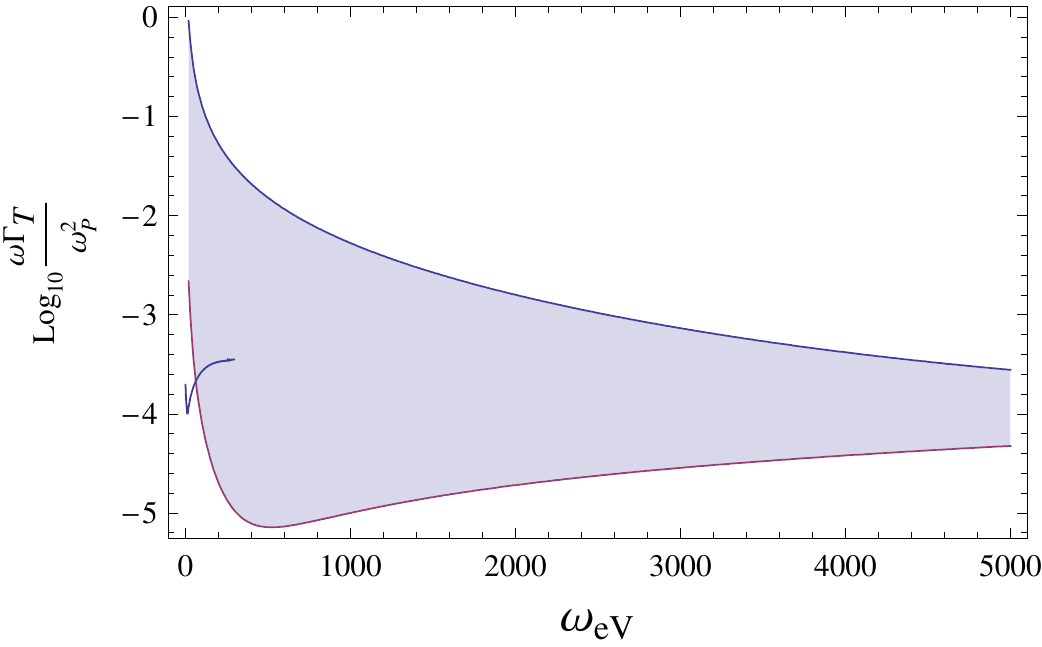}\hspace{1cm}
\caption{The value of $\omega\Gamma/\OP^2$ controls the enhancement of the probability of emission of hidden photons in resonant conditions. 
The shadowed region contains the values for the transverse modes in the whole solar model, with the boundary curves for the Solar center (up) and surface (down). The short line is a lower bound ($\muu=\OP$) for the longitudinal modes for which the relation between the energy and the position at the Sun is fixed. See the text for details.}
\label{GOM}
\end{center}
\end{figure}

In this mass interval, there is always a small region in the Sun where the hidden photon emission is so intensely amplified that outshines the emission from the rest. This is the region where the plasma frequency is tuned to the hidden photon mass $\OP=\muu$, and correspondingly the emission probability is
\begin{equation}
P_{A\rightarrow B}^{\mathrm{res}}=\chi^2\frac{\muu^4}{\omega^2\Gamma_\T^2} \ . 
\end{equation}
The factor $\omega\Gamma_\T/\OP^2$, plotted in Fig.~\ref{GOM}, enhances the mixing and therefore the probability, but if it is too small can invalidate the WM condition. We will find that the energy loss argument imposes values for $\mix$ smaller than $10^{-8}$ in this mass range (See Fig.~\ref{bounds}) so in any case the WM condition is always satisfied.  

If we neglect the small $r$-dependence of $T$ and $\Gamma_\T$ in eq.~\eqref{Tflux} the half-width of this resonance can be easily calculated from eq.~\eqref{PAB} giving
\begin{equation}
\Delta \OP^2= 
2\omega\Gamma_\T \ .  \label{HWT}
\end{equation}
Therefore the resonant emission will take place in a tiny shell of the solar interior, of size $ \Delta r \simeq \Delta \OP^2 (d \OP^2/d r)^{-1}$. The resultant flux at the Earth can be approximated by
\begin{equation}
\frac{d\Phi}{d\omega} \simeq  \Delta r \frac{r^2}{R^2_\oplus}   \frac{\omega\sqrt{\omega^2-\muu^2}}{\pi^2(e^{\frac{\omega}{T}}-1)}\frac{ \muu^4\mix^2}{\omega^2\Gamma_\T^2} \Gamma_\T  = 2\frac{r^2}{R^2_\oplus} \frac{\sqrt{\omega^2-\muu^2}}{\pi^2\frac{d\OP^2}{dr}(e^{\frac{\omega}{T}}-1)} \muu^4 \mix^2  \ \ . \label{flux_2}
\end{equation}
As a remarkable fact, the dependence on $\Gamma_\T$ cancels out. Any improvement in the derivation of $\Gamma_\T$ will not change this result. However, even coming from a tiny shell of the Sun with almost constant temperature $T=T(r(\OP))$, the spectrum of hidden photons does not have a perfect thermal shape; the power of the energy is roughly $\omega$, not $\omega^2$.

Finally the energy loss in hidden photons can be easily calculated. Using $\OP=\muu\ll T$ I find 
\begin{equation}
W_{\gamma'} = 4\pi R^2_\oplus \int_{\muu}^\infty  \frac{d\Phi_\T}{d\omega}\omega d\omega = \frac{16\zeta(3)}{\pi^2} \frac{r^2 T^3}{\frac{d\OP^2}{dr}} \muu^4 \mix^2 \ \ .  \label{W_2}
\end{equation}
Let me remark that while for $\muu\ll 1$ eV the $\muu$-dependence of the hidden photon emission factors out, here in eqs.\eqref{flux_2} and \eqref{W_2} it is implicitly assumed that all the quantities $T,r,\OP,\Gamma_\T,d\OP^2/dr$ are evaluated at the point of the solar model where $\OP=\muu$. These dependences can be read in Fig.~\ref{solarmodel} and will help us to understand the bounds of Sec.~\ref{Sbounds}. 


\subsubsection{Unsuppressed production ($\mathbf{\muu > 295}$ eV)} 

\begin{figure}[t]
\begin{center}
\includegraphics[width=7cm]{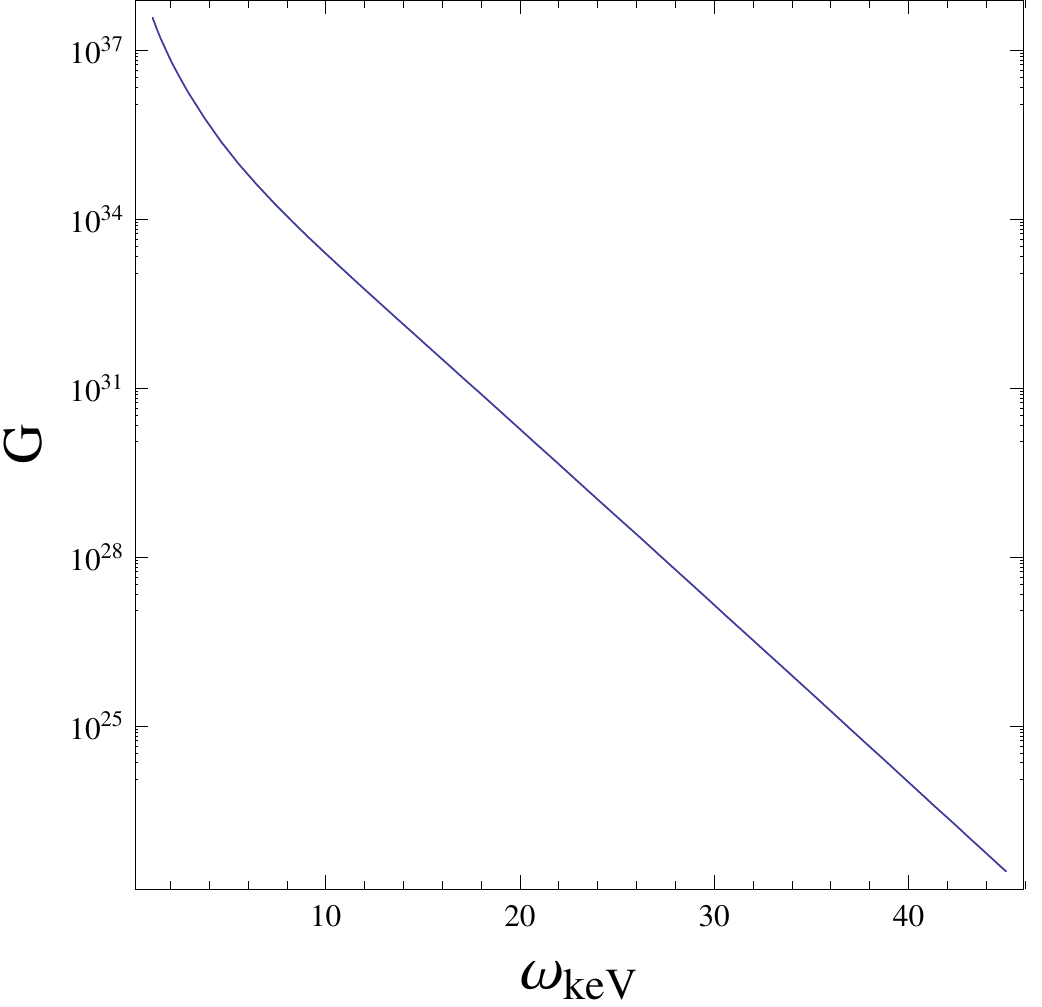}\hspace{1cm}
\caption{The function $G$ gives the flux of $\muu\gg295$ eV hidden photons from the Sun. See the text for details.}
\label{G}
\end{center}
\end{figure}

In this case the emission probability of eq.~\eqref{PAB} is simply $\chi^2$. The flux in eq.~\eqref{Tflux} is therefore independent of $\muu$ if the energy $\omega$ is high enough such that the threshold corrections like the square root are small. In this case we find
\begin{equation}
\frac{d\Phi_\T}{d\omega} \simeq \frac{1}{4\pi R^2_\oplus} \int_0^{R_\odot} 4\pi r^2 d r \frac{1}{\pi^2} \frac{\omega^2}{e^{\frac{\omega}{T}}-1}  \mix^2 \Gamma_\T 
\equiv \frac{\chi^2}{\mathrm{cm}^2\ \mathrm{s}\ \mathrm{eV}}    G(\omega)  \ \ \ ,\label{flux_3} 
\end{equation}
where $G(\omega)$ is a dimensionless function plotted in Fig.~\ref{G}.
In general in this regime most of the production comes from the solar center, where $\Gamma_\T$ and $T$ are bigger and therefore also the possible energies $\omega$. For energies above $\sim 5$ keV Compton scattering dominates over bremsstrahlung and thus we expect an almost Planckian spectrum with a temperature $\sim T_\odot=1.35$ keV, with exponential suppression for $\muu\gg T_\odot$.

\subsection{$B_\0$ production}

Longitudinal plasmons can also resonantly convert into hidden photons. Moreover, since in this case $m_\0^2\simeq\OP^2-k^2$ the resonant conversion will not only happen at a solar shell where $\OP\simeq\muu$ but at \emph{every place in the Sun} where $\OP>\muu$ since $m_\0^2=\muu^2$ is always satisfied by a certain value of $k$. 

As the energy $\omega\simeq\OP$ is almost unrelated with $k$ by eq.~\eqref{dispersion}(L) but $r=r(\OP)$,  it is convenient to approximate the momentum integral in eq.~\eqref{generalflux} by the value at the resonance times the half-width $\Delta k$ (in analogy with eq.~\eqref{HWT} $\Delta k^2$ will be $2\omega\Gamma_\0$) and use 
\begin{equation}
dr = d\OP^2 \left(\frac{d\OP^2}{dr}\right)^{-1}\simeq 2\omega d\omega \left(\frac{d\OP^2}{dr}\right)^{-1} \ , 
\end{equation}
to get
\begin{equation}
\frac{d\Phi_\0}{d\omega} = \frac{r^2}{R^2_\oplus} \frac{\sqrt{\omega^2-\muu^2}}{\pi^2\frac{d\OP^2}{dr}\frac{\omega}{T}}  \muu^4\mix^2   \label{Lflux} \ , 
\end{equation}
which looks exactly the same formula than for the resonant flux of T-hidden photons eq.~\eqref{flux_2} \emph{but} here $r$ and $d\OP^2/dr$ are evaluated at the point where $\OP=\omega$. This limits the $B_\0$ flux to energies $1$eV  $\lesssim \omega\lesssim 295$ eV,  the range of the plasma frequency in the solar model. 

For small $\muu$, again $\muu^4\chi^2$ factors out of the flux and, in analogy with eq.~\eqref{flux_1}, we can define a dimensionless function $F_2(\omega)$ as $(d\Phi_\0/d\omega)($cm$^2$eVs$)(\muu/\mathrm{eV})^{-4}\mix^{-2}$ which is plotted in Fig.~\ref{F}. In this mass regime the $B_\0$ luminosity is 
\begin{equation}
W_T(\muu\ll 1\ \mathrm{eV}) = 1.4\times 10^{43}\ \mix^2 \left(\frac{\muu}{\mathrm{eV}}\right)^4\   \mathrm{Watt} \ \ , \label{W_L}
\end{equation}
which is only slightly smaller than than the $B_\T$ luminosity in eq.~\eqref{W_1}.

Then, while resonant T-hidden photon production proceeds in a tiny shell that emits at all energies, L-hidden photons are resonantly produced in a sphere of radius $r=r(\OP=\muu)$ and the position inside this sphere determines the unique resonant energy $\omega=\OP(r)$.

Unfortunately, beyond $\muu={\OP}_\odot\sim 295$ eV the resonant production is not possible and the emission of $B_\0$'s drops drastically. Note from eq.~\eqref{dispersion} that the L-plasmon energy cannot exceed $\OP$ by much without entering the region where Landau damping is strong, namely $k \gtrsim\sqrt{m_e/T}\OP$. For such large values of $k$ all our approach has to be revised. However, at the view of the bounds that can be extracted with the more powerful resonant emission this seems not to be a fruitful business.

It is easy to show that the WM condition in this case is also satisfied. For $\muu> 1$ eV we will require (at least) $\mix<10^{-8}$ and Fig.~\ref{GOM} shows than at most an enhancement of $10^4$ can be expected. For smaller values of $\muu$ these quantities both scale with $\muu^2$ so the conclusion remains unchanged.
 
\section{Helioscope detection}\label{Shelioscope}

Having addressed the calculation of the hidden photon fluxes at the Earth, in this section I point out that existing axion helioscopes \cite{Sikivie:1983ip,vanBibber:1988ge} like CAST \cite{Zioutas:2004hi,Andriamonje:2007ew} at CERN are indeed capable of detecting these exotic photons. 

\begin{figure}[t]
\begin{center}
\includegraphics[width=17cm]{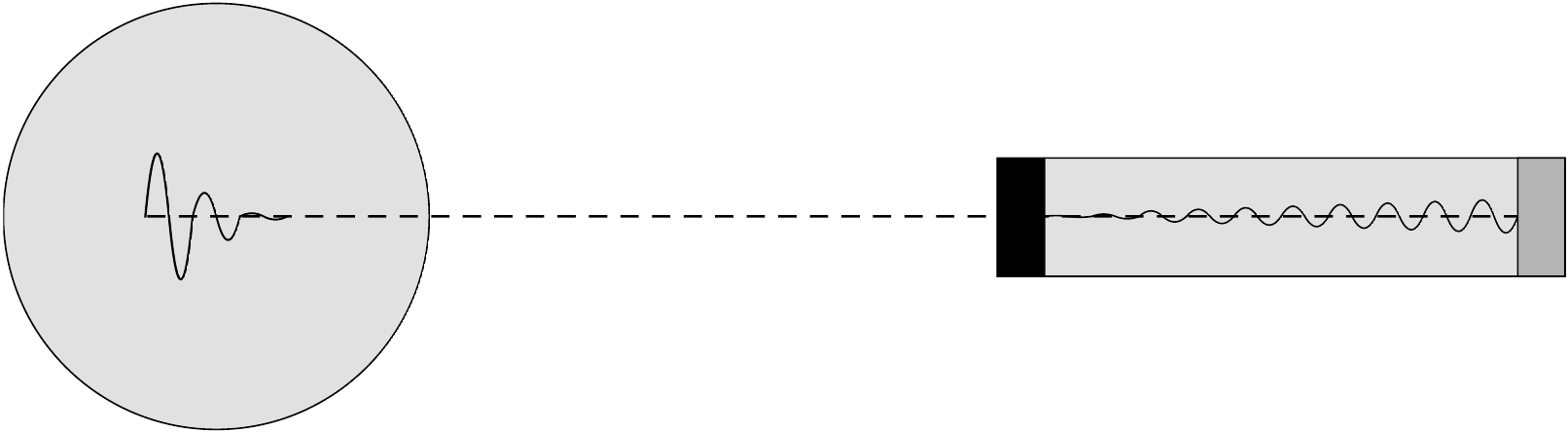}
\Text(-200,75)[c]{\scalebox{1}[1]{$B$}}
\Text(-120,75)[c]{\scalebox{1}[1]{$S\rightarrow A$}}
\Text(-420,100)[c]{\scalebox{1}[1]{$A\rightarrow \tilde S$}}
\Text(-360,50)[c]{\scalebox{1}[1]{$\tilde S \rightarrow S_{R_\odot}\equiv B$}}
\Text(-415,150)[c]{\scalebox{1}[1]{Sun}}
\Text(-90,120)[c]{\scalebox{1}[1]{Helioscope}}
\Text(-170,30)[c]{\scalebox{1}[1]{Shielding}}
\Text(-10,30)[c]{\scalebox{1}[1]{Detector}}
\caption{Schematics of an helioscope experiment looking for hidden photons. Hidden bosons are produced in the Sun's interior from X-ray plasmon conversion and get out almost freely. Only the sterile component ($S$) of $B$ will traverse the helioscope external shielding, leading to the possibility of reconversion into a detectable photon by $S-A$ oscillations. }
\label{helioscope}
\end{center}
\end{figure}

The set up of a typical axion helioscope is depicted in Fig.~\ref{helioscope}. It consists in a (preferably) long cavity pointing towards the Sun where a strong magnetic field is maintained. The cavity is strongly sealed and has a powerful, low background X-ray detector at the end. 

If axions are emitted from the Sun \cite{Dicus:1978fp,Fukugita:1982ep}, they easily pass through the shielding and a few of them can be coherently reconverted into X-rays. This conversion can be understood in terms of axion-photon mixing \cite{Raffelt:1987im}, where the magnetic field acts as the mixing agent, providing the required angular momentum (axions are spin-0 particles). 

In contrast, hidden photons do not need a magnetic field to satisfy angular momentum conservation, they just mix naturally with photons regardless of the presence of the magnetic field, given that $\chi$ and $\muu$ are both non zero. 

The small mixing shift in eq.~\eqref{shift} turns out to be again a correct approximation for the treatment of the photon-hidden photon system both in the Sun-Earth travel and inside the CAST oscillation region. This is because CAST operates either in high vacuum where we can neglect $\pi_a$ (and the WM condition is an exceedingly good approximation), or filled with gas to force resonant conversion, and therefore producing also a non-zero $\Gamma_a$ that cuts off the divergence of the denominators in eq.~\eqref{shift} at reasonable values. The interstellar medium can be also treated as a perfect vacuum.

Note that essentially all the hidden photons emitted from the Sun are propagation states $\propto B_a$ so they do not suffer oscillations. However, these states will be projected into their sterile component $S$ when traversing the helioscope shielding. Such a projection decreases the flux only in a small ${\cal O}(\mix^2)$ factor and can be neglected\footnote{Effects of the atmosphere and further barriers are also at the ${\cal O}(\mix^2)$ level and therefore are also unimportant.}. An initial state $S_a$ traveling through the conversion region (of length $L$) will oscillate into a detectable photon with a probability given by 
\begin{eqnarray}
P_{S_a\rightarrow A_a}=|\langle A_a|S_a(L)\rangle|^2= |\langle A_a(L)|S_a\rangle|^2  &=&\\ 
\frac{\chi^2\muu^4}{(m_a^2-\muu^2)^2+(\omega\Gamma_a)^2}&&\hspace{-1cm}\left(1+e^{-\Gamma_a L}-2e^{-\frac{\Gamma_a L}{2}}\cos{\Delta p_a L}\right) \ , 
\end{eqnarray}
where $\Delta p_a=\sqrt{\omega^2-m_a^2}-\sqrt{\omega^2-\muu^2}$ is the difference in wavenumbers of the photon and hidden photons. In vacuum $\OP=\Gamma=0$ and L-plasmons cannot been excited, so we recover the well-known expression 
\begin{equation}
P_{S_T\rightarrow A_T}=4 \mix^2 \sin^2{\hspace{-2pt}\frac{\Delta p L}{2}} \ \   , \label{vacuum}
\end{equation}
that in the limit $\omega\gg\muu$ leads to 
\begin{equation}
P_{S_\T \rightarrow A_\T}= 4\mix^2 \sin^2{\hspace{-2pt}}\frac{\muu^2 L}{4\omega}  \ \  , \label{lowmass}
\end{equation}
and, for $\muu^2\ll\omega L^{-1}$,  
\begin{equation}
P_{S_\T \rightarrow A_\T}= \frac{\mix^2\muu^4L^2}{4\omega^2} \ \  . \label{lowdistance}
\end{equation}
An interesting situation arises when the hidden photon mass is so large that the argument in the sinus of eq.~\eqref{lowmass} is much larger than one. If the energy resolution of the detector is such that integrates photons with energy between $\omega$ and $\omega+\Delta\omega$ then it effectively integrates several oscillations when $\Delta \omega/\omega \gg 4\pi\omega/(\muu^2L)$. In the limiting case, the sinus is effectively averaged to $1/2$, if the dependence of the flux with the energy is reasonably small. Note that in this case the conversion probability will be effectively  independent of the oscillation path.

If oscillations take place in a medium, even longitudinal excitations could appear. However the detectors used for helioscopes like CAST are designed for the detection of transversely polarized photons. 
For the T-modes a resonant detection is possible if the oscillation volume is filled with a small amount of gas such that $m_\T^2(=\OP^2)=\muu^2$ (above the energy of the highest atomic resonance the dispersion relation of photons in gas is essentially the same than for a plasma). Assuming $\Gamma_ a L\ll 1$, the conversion probability is again independent of the absorption coefficient,
\begin{equation}
P_{S_\T\rightarrow A_\T}= \frac{\chi^2\muu^4L^2}{4\omega^2}  \ \ \ .
\end{equation}
A formula for the range of validity of the WM approximation can be easily provided. For keV photons above the atomic resonances of the gas the Thomson scattering will be the main source of absorption. We find then that $\mix\OP^2/(\omega\Gamma_\T)=\mix3m_e/(2\alpha\omega) \simeq10^6 \mix$(keV$/\omega)\ll 1$ should hold.

In our bounds we should use the values for the most recent and sensitive experiment. The CAST helioscope at CERN has recently published results \cite{Andriamonje:2007ew} from a search of solar axions in a decommissioned LHC magnet of 10 m length in an energy window $0.5$ keV$<\omega< 15$ keV. They found no signal over a subtracted background of $1\sim 4\times10^{-6}$ counts cm$^{-2}$s$^{-1}$keV$^{-1}$ depending on the energy. In order to be conservative I will use a slightly bigger value
\begin{equation}
\Phi_\mathrm{CAST} < 10^{-5} \mathrm{cm}^{-2}\ \mathrm{s}^{-1} \mathrm{keV}^{-1} \ , \hspace{1cm} \mathrm{for} \hspace{1cm} 0.5\  \mathrm{keV}<\omega <15\  \mathrm{keV} \ , 
\label{CAST}
\end{equation}
in the bounds of this paper. This limit arises from the difference of background photons when the helioscope points towards the Sun and when it does not (axions need the magnet to convert into photons). If the hidden photon mass is big enough such that the conversion probability is independent of the length this procedure will subtract the possible signal as well! Therefore, for masses larger than $\sqrt{8\pi\omega^2/(\Delta\omega L)}$ one should use the pure background counts to estimate the bounds. However, one of the 3 detectors of the CAST experiment (the CCD camera) has a X-ray focusing device that in any case increases the flux of photons coming along the magnet direction with respect to those coming from other directions in a sizable amount, making the subtraction harmless. As the exclusion bound derived with this detector is stronger than with the two others (the TPC and the Micromegas), it seems conservative to still use eq. \eqref{CAST} to limit the flux of photons from hidden-sector conversion.  

\section{Bounds and discussion}\label{Sbounds}

\begin{figure}[t!]
\centering
\hspace{-1.3cm}\includegraphics[width=14cm]{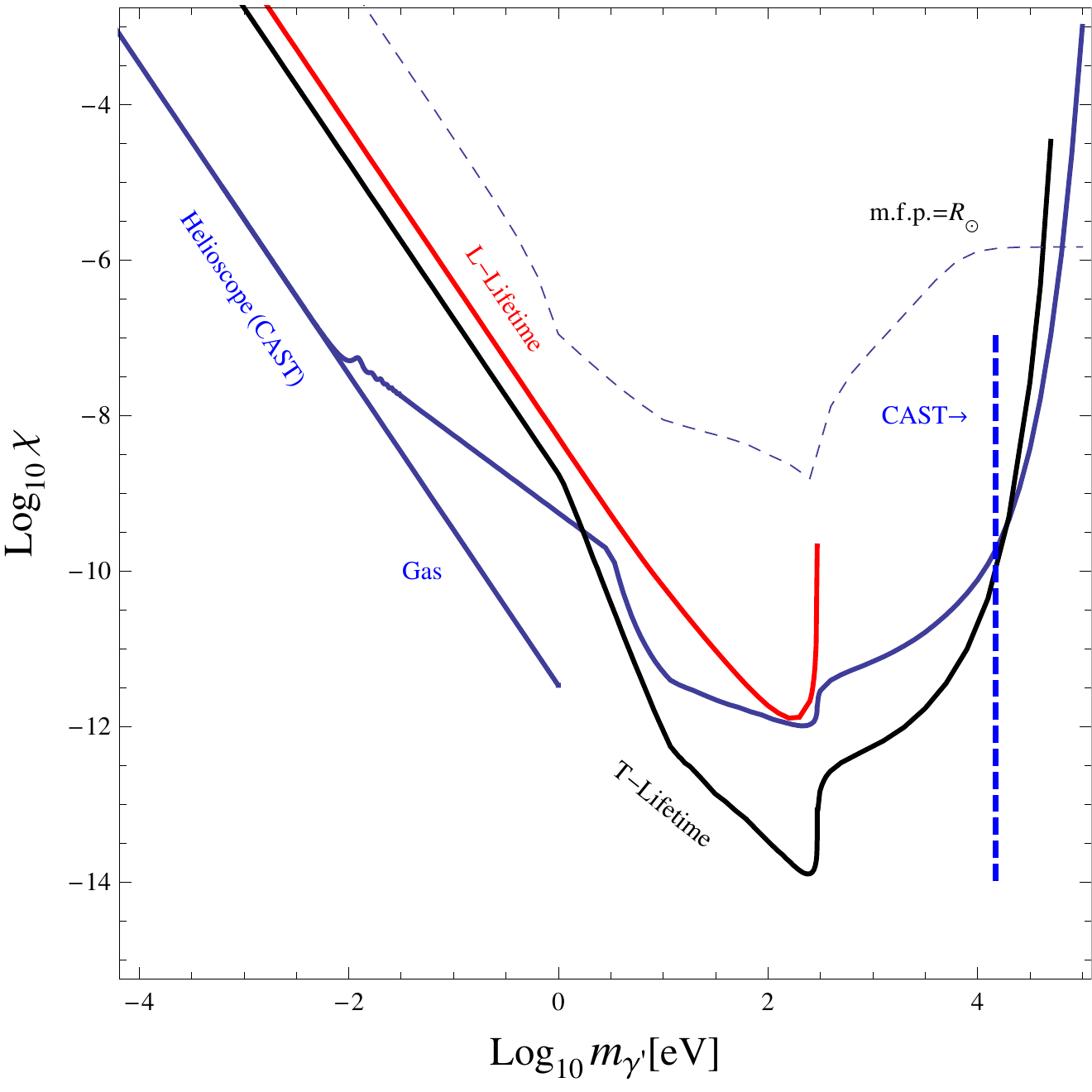}
\vspace{.5cm}
\caption{\label{bounds} Upper limits on the kinetic mixing parameter vs. the mass of the hidden photon from the CAST helioscope and the Solar lifetime argument. See the text for details.}
\end{figure}
%

The energy loss in terms of novel particle species can have dramatical consequences for stellar evolution. 
Hidden photons seem to behave like invisible axions \cite{Dine:1981rt,Kim:1979if,Sikivie:1983ip}, wherever they are produced they leave the Sun without further relevant interactions. Although in principle such a non-standard energy loss could be easily accommodated in a ``present day" solar model by increasing the central temperature a bit over the standard value (the nuclear reaction rates depend strongly on the temperature) the nuclear fuel would be consumed faster and the star would enter sooner in the next stage of stellar evolution \cite{Frieman:1987ui}. 

Theoretical \cite{Frieman:1987ui} and numerical studies \cite{Raffelt:1987yu} incorporating axion loses showed that no present day solar model can be constructed if the exotic luminosity is bigger than the actual solar luminosity $L_\odot$. Under the hypothesis that hidden-photon losses have similar effects and taking $L_\odot=3.83\times 10^{26}$ Watt we can obtain the bounds in Fig.~\ref{bounds} labeled $\T$-Lifetime and $\0$-Lifetime.

The T-Lifetime bound has been obtained by integrating numerically the general expression eq.~\eqref{flux_1} for $B_T$, except for the region of resonant production ($1$ eV$<\muu<295$ eV) for which eq.~\eqref{W_2} has been used. The L-Lifetime bound comes from numerically integrating the energy of the flux in eq.~\eqref{Lflux}. The bound should regard the sum of both T and L contributions but in practice the L contribution does not change the limit appreciably.

Below $\muu\sim 1$ eV the bound is a straight line of slope $-2$ because in this case the hidden photon flux is just proportional to $\mix^2\muu^4$. In the range $1$ eV $\lesssim \muu\lesssim 295$ eV the resonant production dominates and improves the bound down to $\mix\sim 10^{-14}$ where the resonance takes place in the solar core. 
The change of slope of the T-Lifetime bound around $\muu\sim 10$ eV can be traced back to the temperature dependence of eq.~\eqref{W_1}, as can be seen in Fig.~\ref{solarmodel} since in this regime $\OP$ corresponds to $\muu$. There we can also see that the inflection point at $\sim 290$ eV it is due to the decrease of the volume of emission through the factor $r^2$ of eq.~\eqref{W_1}. 

Beyond $295$ eV the T-Lifetime bound sharply worsens until it reaches the bound due to the solar bulk emission. For $\muu\gg 10$ keV the bound vanishes because of the lack of photons with enough energy.

Let us now turn our attention to the  CAST bounds coming from the non-observation of X-ray photons in the energy window $0.5-15$ keV. We have already discussed that CAST is not sensitive to $B_\0$'s so we restrict the discussion to the non-observation of transverse excitations.

The hidden photon flux is a decreasing function of the energy since it comes essentially from bremsstrahlung (see also eq.~\eqref{flux_2} and Figs.~\ref{F} and \ref{G}) so we can restrict the bounds to the most interesting  (yet reasonably sized) interval, which is $1$ keV above $\muu$.

Integrating the $B_\T$ flux in eq.~\eqref{Tflux}, or eq.~\eqref{flux_2} when resonance dominates, times the vacuum conversion probability eq.~\eqref{vacuum} in the mentioned energy interval and using eq.~\eqref{CAST} I get the bound labeled ``Helioscope (CAST)" in Fig.~\ref{bounds}. 

The onset of the oscillation regime in the CAST detector is clearly seen at $\muu\sim\sqrt{4\times 1\mathrm{keV} /10\ \mathrm{m}}\sim 0.01$ eV. Below this mass, $S-A$ oscillations are never complete and eq.~\eqref{lowdistance} holds. The CAST signal is then proportional to $\mix^4\muu^8$, and therefore the slope of the Helioscope line is again $-2$.
Soon above this mass, the oscillation length in CAST is so small that the squared sine of eq.~\eqref{vacuum} gives a factor of $1/2$ when we average over energy. In this regime, but below $1$ eV, the CAST signal is proportional to $\mix^4\muu^4$ and therefore the Helioscope line has slope $1$. In order to mantain this slope down to $10^{-4}$ eV the CAST oscillation length should be increased in a factor $10^4$ (or the energies detected lowered in a similar factor), however the region of improvement is already excluded by laser experiments \cite{Ahlers:2007qf}.

Assuming that the same bound eq.~\eqref{CAST} holds even with gas filling the oscillation region, and varying the plasma frequency of the gas in the reasonable range $0.01-1$ eV, the bound labeled ``Gas" could be achieved. While it is not clear if this procedure is realistic for higher masses, Fig.~\ref{bounds} shows that it would be extremely interesting.

In the mass range $\sim 1-10^4$ eV, the CAST bound lies in a region ruled out by the energy loss argument. Since the emission has been calculated assuming an unperturbed solar model we find that the CAST bound is not consistent in this mass range. The observation of a  hidden photon flux corresponding to this region would imply drastic changes in our current picture of solar structure. 

Note that the curve scales with the fourth power of the CAST limit eq.~\eqref{CAST} so a huge improvement in sensitivity would be needed to beat the energy-loss bound in this range. Observe however, that there is no change of slope in the Helioscope line at $\mu\sim 1$ eV where the T-Lifetime starts to be dominated by resonant emission. The reason is that the resonant production starts at $\OP\sim 1$ eV, close to the solar surface where the temperature is much smaller than the CAST energy window. Therefore, some hope of improvement in the mass range $1-10$ eV relies in lowering the lower threshold until ${\cal O}(\mathrm{eV})$ energies.

The top of the energy window, $15$ keV, limits the range of hidden photon masses that can be testable with CAST. The dashed vertical line shows this limitation. Still I have continued the ``Helioscope" line up to somehow higher masses by assuming the same sensibility eq.~\eqref{CAST} and no threshold. Interestingly enough, this shows that for $\muu>25$ keV the CAST bound \emph{would be} again consistent with, and more powerful than, the energy loss bound. Unfortunately for $\muu>40$ keV the hidden photon flux cannot be calculated from an unperturbed solar model and the bounds become indeterminate. 

Finally, it is interesting to compare these results with the earlier works \cite{popov:1991,Popov1999}. The energy loss bounds presented here are weaker except in the resonant region, whose effects were apparently not considered before. The extremely good results of the CAST collaboration, are however able to reverse the situation for low masses. Even with the lower flux calculated in this paper the CAST exclusion line goes deeper, not only than the former helioscope bound, but also than the energy loss limit. Finally it is interesting to note that the lifetime bound derived here extends one order of magnitude further in mass. Therefore, except for two islands around $\muu\sim 1$ eV and $\muu\sim 1$ keV, the bounds presented here improve the earlier limits.

Our last check should be to ensure that the $\tilde S$ absorption inside the Sun is negligible as it has been assumed in all the above. If this were not the case, not only the hidden photon flux will decrease with respect to the above estimates but the same solar internal structure will require an unacceptable readjustment because of the resulting non-local energy transfer \cite{Raffelt:1988rx}. 
I have checked that absorption is not significant except for the case of massive hidden photons $\muu\gg 295$ eV with $\mix\gg10^{-6}$. The main reasons for that are that absorption in the low mass range is suppressed by $\muu^4$ and the resonant regions in the intermediate mass regime are never too wide. For illustration purposes the line labeled ``m.f.p.$=R_\odot$" shows the value of the mixing parameter $\mix$ for which a T-hidden photon produced in the solar center would have a mean free path of the order of the solar radius. This mean free path is an average over the radial trajectory using at every place of the Sun the highest possible value of the absorption rate (energies near the plasma frequency), so it is again a very conservative estimate.

In this context I should emphasize that the region above the ``Lifetime" and ``Helioscope" curves is not strictly ruled out by the energy loss and CAST limits since it assumes that hidden photon emission is a small perturbation of the standard solar model \cite{Bahcall:2004pz}. Only adding the fact that the standard solar model agrees very well with helioseismological data and the observed neutrino fluxes, and these are typically very sensitive to the internal structure of the Sun, one concludes by ``reductio ad absurdum" that this region is severely excluded.  

As a final remark we should keep in mind that these bounds can be completely different if, in addition to the hidden photon, other low mass  particles exist in the hidden sector. In particular, the stellar emission of particles of mass $\lesssim 10$ keV, charged under $U(1)_\mathrm{h}$ does not vanish in the limit $\muu\rightarrow 0$ \cite{Davidson:1993sj} (actually this can be even true if these particles are much heavier \cite{Hoffmann:1987et}). In this case low hidden photon masses are constrained as much as masses of the order of the stellar temperature.

\section{Conclusions}\label{Sconclusions}

I have addressed the calculation of the solar emission of a hypothetical hidden sector photon $B^\mu$ mixing kinetically with the standard model ordinary photon. I have shown that a resonant effect is possible when the dispersion relation of solar plasmons fits the particle-like dispersion relation of the hidden photon. This happens for transverse plasmons if the hidden photon mass $\muu$ lies in the range $1\sim295$ eV (the range of the plasma frequency in the solar model used) and for longitudinal plasmons as long as $\muu\lesssim295$ eV. 

The conservative requirement that the hidden photon luminosity should not exceed the solar standard luminosity bounds the amount of kinetic mixing up to $\mix\lesssim10^{-14}$ depending on the mass and the polarization. At masses beyond $1$ eV, where the strongest bound is reached, the emission of transversally polarized hidden photons dominates over the emission of longitudinal ones. Below this mass both polarizations contribute in a similar amount.

At low masses the bounds are weaker, relaxing proportionally to $\muu^2$.
However, the non observation of a signal in the CAST axion helioscope improves the bounds in this region up to 2 orders of magnitude. Altogether, these are the best limits on the mixing parameter $\mix$ in the range $3$ meV$<\muu<40$ keV.   A small room for improvement is available for large masses $\muu\gg 10$ keV if the CAST detectors were to rise their top energy threshold.

It should be interesting to extend this study to other stellar objects like supernovae, white dwarfs, red giants and horizontal branch stars, since these can provide stronger bounds than the Sun, specially at masses where a resonant production is possible. 

\section*{Acknowledgements}

I would like to thank Andreas Ringwald for nice discussions and for reading this manuscript, Igor Irastorza and Jaime Ruz for interesting remarks about the CAST helioscope, Joerg Jaeckel and Holger Gies for a useful comment concerning the CAST bounds and Konstantin Zioutas for inviting me to present this work to the CAST collaboration members, to whom I also acknowledge fruitful conversations and a warm welcome. 
\vspace{1cm}

\appendix
\noindent
{\LARGE \bf Appendix}
\section{Thomson dispersion of longitudinal plasmons}\label{A}
\nonumber\notag

The dispersion relation for longitudinal plasmons allows for space-like excitations. These plasmons can be coherently absorbed by the electrons in the plasma giving rise to an order ${\cal O}(\alpha)$ absorption - the so-called Landau damping \cite{Landau:1946,tsytovich:1961}- which can be important for the bulk emission. However, only time-like L plasmons, whose ``Landau" absorption is kinematically forbidden, can be resonantly produced making natural to restrict this discussion to them. The next order contribution to the absorption rate of plasmons is due to Thomson dispersion, $\gamma_a e^-\rightarrow \gamma_{a'} e^-$.  Given that typical energies in the solar interior are much smaller than the electron mass, it is justified to focus on the non-relativistic limit $m_e\rightarrow\infty$. 
It is worth noting that in this case the transitions to transverse plasmons $\gamma_\0 e^-\rightarrow \gamma_\T e^-$ are suppressed by phase space since $\omega_\T(k)$ grows much faster with $k$ than $\omega_\0(k)$ and the target electrons cannot transfer energy efficiently at such small energies. Therefore I only address the calculation of the $\gamma_\0 e^-\rightarrow \gamma_\0 e^-$ process.

Let us begin by defining the polarization vectors for the initial and final photon states.  Writing $k^\mu=(\omega;0,0,k)$ f and $k'^\mu=(\omega';k'\sin\theta,0,k'\cos\theta)$ for the initial and final photon 4-momenta we can choose
\begin{equation}
\epsilon_\0^\mu \frac{1}{\sqrt{\omega^2-k^2}}(k;0,0,\omega)\ \ ; \ \ {\epsilon '}_{L}^\mu=\frac{1}{\sqrt{\omega'^2-k'^2}}(k';\omega'\sin\theta,0,\omega'\cos\theta)  \  . 
\end{equation}
The matrix element can be written as
\begin{eqnarray}
{\cal M}= {\epsilon'}^*_{\mu} {\epsilon}_\nu  {\cal M}^{\mu\nu} \ , 
\end{eqnarray}
where 
\begin{eqnarray}
{\cal M}^{\mu\nu}=-e^2\sqrt{Z}\sqrt{Z'}\ \overline{u}(p',s')
\left[\frac{\gamma^\mu\slashed{k}\gamma^\nu+2 \gamma^\nu p^\mu}{2p\cdot k+k_\mu^2}+
\frac{\gamma^\nu\slashed{k}'\gamma^\mu - 2 \gamma^\mu p^\nu}{2p\cdot k'-k_\mu^{\prime2}}\right]u(p,s) \ , 
\label{M}
\end{eqnarray}
and $Z,Z'$ are the wave function renormalization factors of the initial and final plasmons given by 
\begin{equation}
Z_\0(k)\equiv \widetilde{Z}_\0 \frac{\omega^2}{\omega^2-k^2}=\frac{2(\omega^2-v_*^2k^2)}{3\OP^2-2(\omega^2-v_*^2k^2)}\frac{\omega^2}{\omega^2-k^2} \ , 
\end{equation}
where $v^2_*=5T/m_e\ll 1$ and $\omega,k$ are understood to satisfy the dispersion relation eq.~\eqref{dispersion}($\0$). In practice, because we restrict ourselves to time-like plasmons $\omega\gtrsim k$, we can set $\widetilde Z_\0\sim 1$.

Because of charge conservation, $\cal M^{\mu\nu}$ satisfies necessarily the conditions
\begin{eqnarray}
k_\nu {\cal M}^{\mu\nu} =\omega {\cal M}^{\mu 0}-k {\cal M}^{\mu 3} = 0    \ \ \ ; \ \ \ 
k'_\mu {\cal M}^{\mu\nu} =\omega' {\cal M}^{0 \nu}-k'^i {\cal M}^{i \nu} = 0  \ ,\end{eqnarray}
that can be used to express ${\cal M}$ in a very convenient way
\begin{eqnarray}
{\cal M}_{L}&=&\frac{\sqrt{(\omega^2-k^2)(\omega'^2-k'^2)}}{\omega\omega'}\left({\cal M}^{13}\sin\theta +{\cal M}^{33}\cos\theta \right)  \label{M33}  \ .
\end{eqnarray}
Note that the renormalization factors in eq.~\eqref{M} will cancel the prefactor in the above equation.

In the non relativistic limit we can use $p^\mu= {p'}^\mu=(m;\mathbf{0})$ for the Dirac spinors and we can neglect the squared 4-momenta of the photons in the denominators of eq.~\eqref{M}. Now it is easy to evaluate ${\cal M}^{i3}$. Recall that $\overline{u}(s')\gamma^i u(s)$ and $p^i$ are proportional to the velocity of the electrons and therefore are suppressed with respect to $\overline{u}(s')\gamma^0 u(s) = u^\dagger (s') u(s) = 2 m {\xi'}^\dagger\xi$. This argument is sufficient to neglect the terms proportional to $p^\mu$ in the numerators. Moreover, using the commutation relations $\{\gamma^\mu,\gamma^\mu\}=2g^{\mu\nu}$ and also $(\gamma^0)^2=-(\gamma^i)^2=1$, it turns out that terms containing three gamma matrices can be either reduced to terms containing only one or they appear in pairs that cancel out. Finally, ${\cal M}^{13}$ is found to be negligible compared with ${\cal M}^{33}\simeq 2e^2 {\xi'}^\dagger\xi$ and the properly averaged squared matrix element is
\begin{equation}
|{\cal \widetilde M}|^2 \equiv \frac{1}{2}\sum_{s,s'} |{\cal M}|^2=
4e^4\cos^2{\hspace{-2pt}\theta} \ .
\end{equation}
The averaged dispersion rate is obtained by integrating the Lorentz invariant phase space of the final particles and averaging over the thermal distribution of electrons in the initial state, weighting with the appropriated stimulation/blocking factors. In practice, however, given that electrons are non-relativistic we can avoid their thermal average and blocking factors and simply multiply by the electron density $n_e$ to get 
\begin{eqnarray}
\Gamma_\0^\mathrm{A}&=& \frac{n_e}{2\omega2m_e}  \int\frac{d^3k'}{(2\pi)^32\omega'}\frac{d^3p'}{(2\pi)^32E'_e}(2\pi)^4\delta^4(p'+k'-p-k)|{\cal \widetilde M}|^2(1+n_\mathrm{BE}(\omega'))  \\
&=& \frac{\alpha^2 n_e}{\omega m_e}  \int\frac{2\pi dk'{k'}^2d\cos\theta}{\omega'm_e}\delta(E'+\omega'-m-\omega)\cos^2{\hspace{-2pt}\theta}(1+n_\mathrm{BE}(\omega'))  \\
&=& \frac{\alpha^2 n_e}{\omega m_e}  \int^k_{k'_m}\frac{2\pi dk'{k'}^2}{\omega'E'}\frac{\cos^2{\hspace{-2pt}\theta}}{\frac{k k'}{E'}}(1+n_\mathrm{BE}(\omega'))=
                        \frac{8\pi\alpha^2 n_e}{9 m_e T} \frac{k}{\OP} (1+n_\mathrm{BE}(\omega')) \ , 
\end{eqnarray}
where in the second step I have used the $\cos\theta$-dependence of the energy of the outgoing electron 
$E'=\sqrt{m^2+(\mathbf{k-k'})^2}=\sqrt{m^2+k^2+{k'}^2-2kk'\cos\theta}$ to remove the Dirac's delta enforcing energy conservation. Together with the dispersion relation, this establishes the relation
\begin{equation}
\frac{k'}{k}= \frac{3T+\OP\cos\theta}{3T+\OP} \ , 
\end{equation}
that bounds the integral over $k'$ to the small interval $(k'_m=k(1-2\OP/3T),k)$. Finally note that the stimulation factor $1+n_\mathrm{BE}(\omega')$ cancels the term in the parenthesis of eq.~\eqref{weldon} since $\omega\simeq\omega'$ so we obtain
\begin{equation}
\mathrm{Im}\{ \pi_\0\} = -\omega \frac{8\pi\alpha^2 n_e}{9 m_e T} \frac{k}{\OP} \ . 
\end{equation}

\bibliographystyle{h-physrev3}


\end{document}